\documentclass{vietnam}

\newcounter{daggerfootnote}

\def\be{\begin{equation}}
\def\ee{\end{equation}}
\def\bea{\begin{eqnarray}}
\def\eea{\end{eqnarray}}

\newcommand{\Photo}{}

\begin{document}
\title{CosmiXs: Improved spectra for dark matter indirect detection \footnote{\sl To appear in the proceedings of the 29th International Symposium on Particles, Strings and
Cosmology (PASCOS 2024), Quy Nhon, Vietnam, 7-13 July, 2024.}}

\author{Chiara Arina \footnote{Electronic address: chiara.arina@uclouvain.be}}
\address{Centre for Cosmology, Particle Physics and Phenomenology (CP3), Universit\'e catholique de Louvain, Chemin du Cyclotron 2, 1348 Louvain-la-Neuve, Belgium}
\author{Mattia Di Mauro \footnote{Electronic address: dimauro.mattia@gmail.com}}
\address{Instituto Nazionale di Fisica Nucleare, Sezione di Torino, Via P. Giuria 1, 10125 Torino, Italy}
\author{Nicolao Fornengo \footnote{Electronic address: nicolao.fornengo@unito.it}}
\address{University of Torino and Istituto Nazionale di Fisica Nucleare, Sezione di Torino, via P. Giuria 1, 10125 Torino, Italy}
\address{Department of Physics, University of Torino, Via P. Giuria 1, 10125 Torino, Italy}
\author{Jan Heisig \footnote{Electronic address: heisig@physik.rwth-aachen.de}}
\address{Institute for Theoretical Particle Physics and Cosmology, RWTH Aachen University, D-52056 Aachen, Germany}
\author{Adil Jueid \footnote{Speaker, Electronic address: adiljueid@ibs.re.kr}}
\address{Particle Theory and Cosmology Group, Center for Theoretical Physics of the Universe, Institute for Basic Science (IBS), Daejeon, 34126, Republic of Korea}
\author{Roberto Ruiz de Austri \footnote{Electronic address: rruiz@ific.uv.es}}
\address{Instituto de F\'{\i}sica Corpuscular, CSIC-Universitat de Val\`encia, E-46980 Paterna, Valencia, Spain}

\maketitle
\abstracts{The spectra of stable particles produced from dark matter (DM) are one of the most important ingredients to calculate the fluxes for DM indirect detection experiments. At energies above a few GeV, most of the particles are produced following a complex sequence of phenomena including resonance decays, QED and QCD final-state radiation, radiation of weak gauge bosons, hadronization and hadron decays. 
In this contribution, we discuss improvements on the calculation of the energy spectra at the source using state-of-the-art tools that include effects that were not taken previously into account. We include helicity information of the particles produced in the annihilation channels, which leads to proper inclusion of electroweak radiation during the entire showering history. These effects are taken into account using the \textsc{Vincia}, which is based on the helicity-dependent antenna shower formalism. Off-shell contributions are also taken into account for annihilation channels into $WW$ and $ZZ$ through the four-body processes into fermions and for DM masses below the gauge boson mass. We also revisit the tune of the Lund fragmentation function parameters in \textsc{Pythia} using LEP data at the $Z$-boson pole. The spectra of cosmic messengers are provided for DM masses between 5 GeV and 100 GeV and are publicly distributed this \href{https://github.com/ajueid/CosmiXs.git}{GitHub repository}.}

\section{Introduction}
\label{sec:intro}

Dark matter (DM) remains one of the most significant mysteries in physics albeit decades of theoretical and experimental searches. Indirect detection is still one of the promising discovery methods of DM signals by seeking excesses in the fluxes of cosmic messenger particles such as positrons ($e^+$), antiprotons ($\bar{p}$), gamma rays ($\gamma$), neutrinos ($\nu$) and antinuclei ($\overline{\rm D}, {}^3\overline{\rm He}$), which are products of DM annihilation into Standard Model (SM) particles followed by their decays.\cite{Gaskins:2016cha} Weakly Interacting Massive Particles (WIMPs) are particularly compelling DM candidates, emerging naturally in several Beyond the Standard Model (BSM) frameworks, including supersymmetry. Experiments such as AMS-02, {\it Fermi}-LAT, SuperK, and IceCube are actively searching for DM-induced signals in the GeV-TeV energy range.\cite{Leane:2020liq} 

The theoretical modeling of the fluxes of cosmic messengers strongly depends on the energy spectra at the source which is usually calculated using multi-purpose Monte Carlo event generators such as \textsc{Pythia}. Poor Particle Physics Cookbook for DM indirect detection (labeled \textsc{PPPC} hereafter)\,\cite{Cirelli:2010xx} is considered to be the standard tool for this task.
More recently, the \textsc{HDMSpectra}\,\cite{Bauer:2020jay} improved the results of the \textsc{PPPC} especially for heavy DM masses by using analytical calculations of all the fragmentation processes in the unbroken SM phase. However, issues arise from matching the results at the electroweak scale. Furthermore, \textsc{QCDUnc}\,\cite{Amoroso:2018qga,Jueid:2022qjg,Jueid:2023vrb} has contributed to the improvements of the hadronization parameter tuning in \textsc{Pythia}~8 by fitting the parameters of the fragmentation using LEP and SLD data and furthermore estimated the QCD uncertainties resulting from both non-perturbative (hadronization) and perturbative (parton showers) physics.

In this study, we aim to improve the theoretical predictions of the spectra of cosmic messengers at the source by using \textsc{Vincia}\,\cite{Fischer:2016vfv} based on the helicity-dependent antenna shower formalism and using \textsc{Pythia}~8.3 for the rest of the physics. The calculation of the parton-level matrix elements is done using \textsc{MadDM}\,\cite{Ambrogi:2018jqj} which provides the output in the LHE format and includes all the spin information of the produced particles thereby properly incorporating the electroweak corrections in our predictions. Compared to the previous results in the literature, our work provides enhanced accuracy for DM masses above $5$ GeV. The spectra are publicly available in this \href{https://github.com/ajueid/CosmiXs.git}{GitHub repository}\,\cite{Arina:2023eic,DiMauro:2024kml} and can be applied to various DM scenarios including the case of a decaying DM through a proper rescaling of the spectra. The rest of the manuscript is organized as follows. In section \ref{sec:physics} we discuss the physics modeling and summarize the main novelties of our analysis including a brief mention of the tuning results. The comparison of our results with those of the \textsc{PPPC}, \textsc{HDMSpectra} and \textsc{QCDUnc} is provided in section \ref{sec:results}. We conclude in section \ref{sec:conclusions}.

\section{Physics modeling}
\label{sec:physics}

Following DM annihilation or decay processes into intermediate states such as leptons, quarks, gluons or massive bosons, a number of phenomena leads the production of stable states at production such as photons, positrons, neutrinos and antiprotons. Additional radiation including QED emissions of photons and charged particles, QCD emissions of gluons and quarks and electroweak boson radiation leads to complex kinematical features. For this, the use of Monte Carlo event generators is necessary to properly model the differential energy distributions. Once reaching an energy scale below ${\cal O}(1)$ GeV, the process of hadronization starts to take over which leads to the confinement of colored particles into color-neutral hadrons. In \textsc{Pythia}, the hadronization is modeled using the string model  with a left-right symmetric function which under minimal assumptions depends on four parameters. The hadronization process has some intrinsic uncertainties that range  from $10\%$ to about $50\%$ depending on the kinematical region, the DM mass and annihilation channel.\cite{Jueid:2022qjg,Jueid:2023vrb} 

The calculation of the energy spectra at the source, by \textsc{PPPC} and \textsc{QCDUnc}, relies on the resonance approach which assume that DM annihilation occurs through the production of a spinless particle (${\cal R}$) in $e^+ e^-$ collisions and $E_{\rm CM} = 2 m_\chi$ which then decays isotropically into a pair of SM particles. This method has many limitations including its inability to account for polarization effects and off-shell contributions. Therefore, these shortcomings lead to inaccurate predictions for the DM spectra especially for annihilation channels involving gauge bosons and SM leptons. To address these issues, we use \textsc{MadGraph\_aMC@NLO} and \textsc{MadDM} to generate the matrix elements for the annihilation channels under consideration thus preserving spin and polarization information which leads to proper inclusion of electroweak corrections. Below we highlight the main novelties of our analysis:

\begin{itemize}
    \item We use \textsc{MadDM} which we interface to \textsc{Pythia} 8 together with \textsc{Vincia} shower plugin being the default option. \textsc{MadDM} produces LHEF where helicity information is provided.
    \item We calculate the spectra for two new annihilation channels: $\chi\chi \to HZ,\gamma Z$.
    \item For the case of $WW/ZZ$ annihilation channels we generate the spectra of the four-body decays and DM masses down to 5 GeV.
    \item For one-loop induced annihilation channels ($gg,\gamma\gamma,\gamma Z$), we take into account the full one-loop effects instead of effective couplings.
    \item We carry out a new tuning of the hadronization model parameters using a set of measurements performed at the Z-boson pole. 
\end{itemize}

\begin{figure}[!t]
\centering
\includegraphics[width=0.49\linewidth]{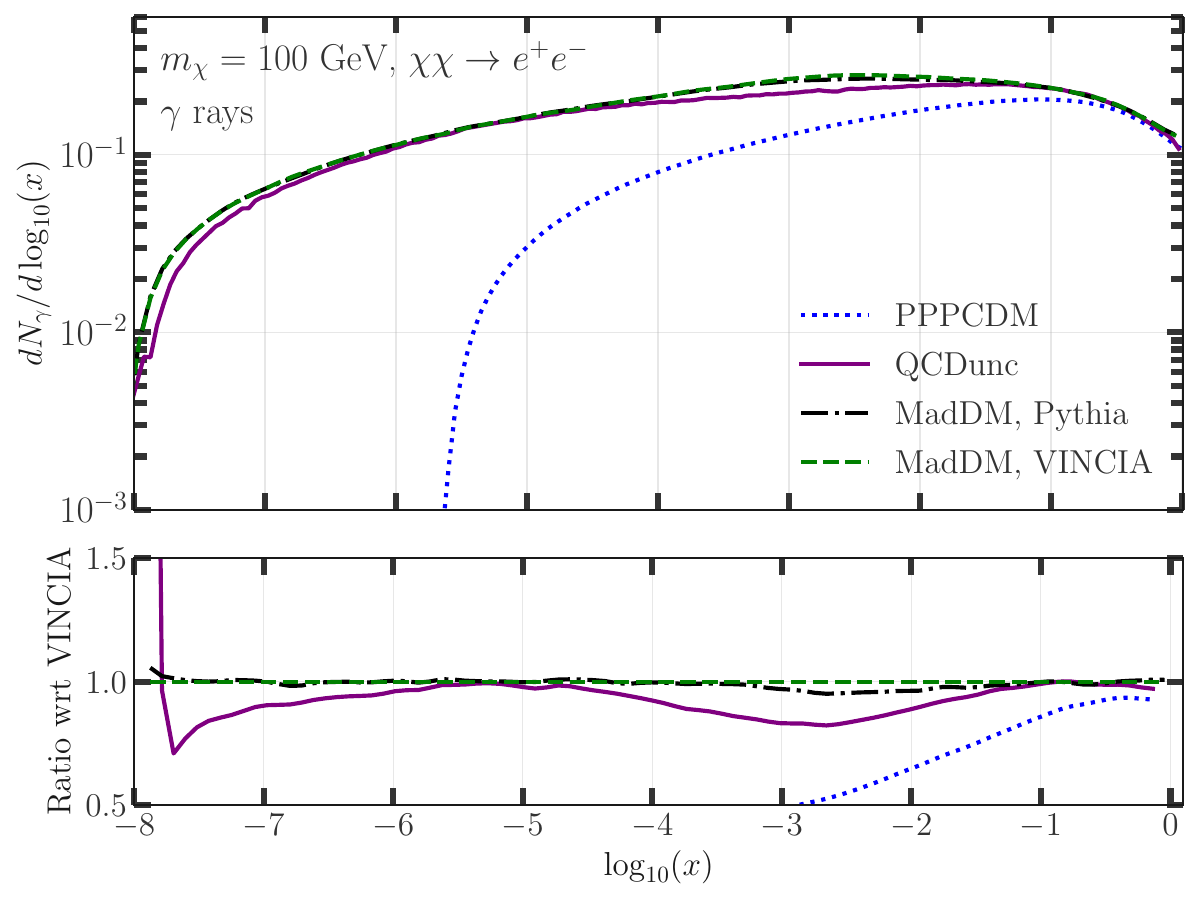}
\includegraphics[width=0.49\linewidth]{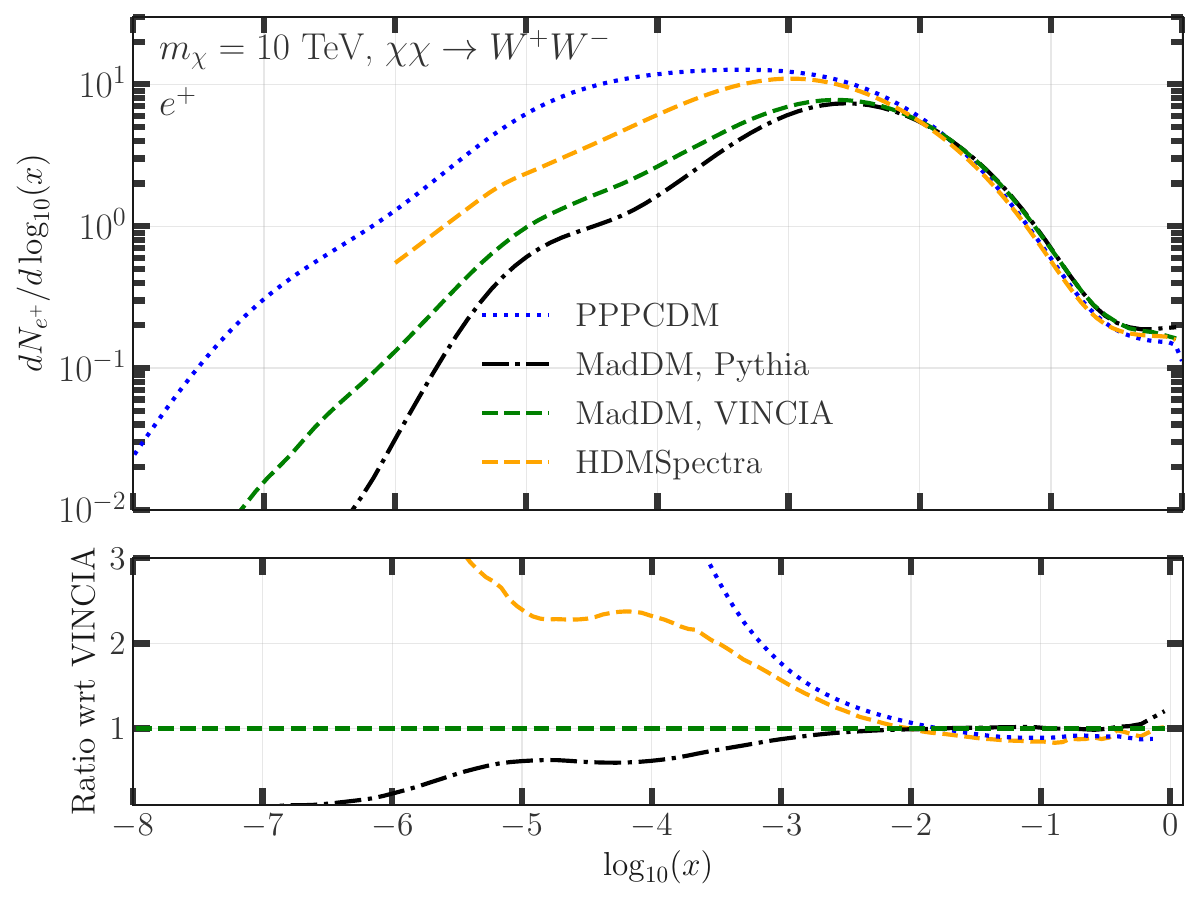}
\caption{Comparison among the spectra obtained with our analysis using the \textsc{Pythia} and \textsc{Vincia} shower algorithms with the results of \textsc{PPPC}, \textsc{HDMSpectra} and \textsc{QCDUnc}. Here we show the spectrum of $\gamma$-rays in $e^+ e^-$ annihilation and $m_\chi = 100$ GeV (left) and the spectrum of positrons for $W^+W^-$ annihilation and $m_\chi = 10$ TeV (right).}
\label{fig:comparison}
\end{figure}

\section{Results}
\label{sec:results}

We briefly discuss some of the results in this section and we refer the interested reader to the main manuscript for more details.\cite{Arina:2023eic} Fig. \ref{fig:comparison} shows a comparison with \textsc{PPPC}, \textsc{HDMSpectra} and \textsc{QCDUnc} for the spectra of $\gamma$-rays and $m_\chi = 100$ GeV (left panel) and the spectra of positrons and $m_\chi = 10$ TeV (right panel). For the spectra of $\gamma$-rays in the $e^+ e^-$ channel with DM masses below 1 TeV, the \textsc{PPPC} underestimates the yields at low energies as introduces a higher cut-off parameter which characterize the transverse momentum threshold for photon emissions off a lepton line. The agreement with \textsc{QCDUnc} is generally good except in the intermediate-energy regions, disagreement are up to $20\%$,  where the role of EWBR becomes quite important. For DM masses above 1 TeV and annihilation into leptons and massive bosons, the effects of EWBR dominate over the other effects and lead to even bigger differences. For instance, the improved handling of the helicity of the produced particles and the inclusion of all the possible trilinear interactions in \textsc{Vincia} leads to an enhancement of the accuracy over the results of the \textsc{PPPC} which adds electroweak corrections by brute force. While \textsc{HDMSpectra} incorporates similar electroweak corrections, deviations in the low-energy part of the spectrum may be attributed to the issues in the matching and soft interference treatments (see the right panel of Fig. \ref{fig:comparison}).

\section{Conclusions}
\label{sec:conclusions}

We presented improved predictions of particle spectra from DM annihilation/decay for various cosmic messengers including $\gamma$-rays, positrons, antiprotons, and neutrinos using the helicity-dependent \textsc{Vincia} shower algorithm which is now a plugin of \textsc{Pythia} 8. 
This approach incorporates new effects that were not fully taken into account in previous results, including all the trilinear boson interactions, full soft-coherence in final-state radiations, and resummation of electroweak corrections through Sudakov. 
We have also calculated the spectra for two new annihilation channels ($\gamma Z$ and $HZ$) and included full off-shell contributions for DM annihilation into massive gauge bosons below their production threshold. We also improved on the tuning of the \textsc{Pythia} 8 parameters for the Lund fragmentation function using LEP data at the $Z$-pole. Our results lead to a precision below or around $10\%$ in the energy regions critical for DM indirect detection experiments and provide vital input for LHAASO, HAWC, CTA and AMS experiments. The results are publicly available on this \href{https://github.com/ajueid/CosmiXs.git}{GitHub repository}.

\section*{Acknowledgments}

N.F.~and M.D.M.~acknowledge support from the Research grant {\sl TAsP (Theoretical Astroparticle Physics)} funded by Istituto Nazionale di Fisica Nucleare (INFN). The work of A.J. is supported by the Institute for Basic Science (IBS) under the project code, IBS-R018-D1. C.A. acknowledges support by the F.R.S.-FNRS under the “Excellence of Science” EOS be.h project no. 30820817. J.H.~acknowledges support by the Alexander von Humboldt foundation via the Feodor Lynen Research Fellowship for Experienced Researchers and
Feodor Lynen Return Fellowship. R.R. acknowledges support from the Ministerio de Ciencia e Innovación (PID2020-113644GB-I00) and the GVA Research Project {\sl Sabor y Origen de la Materia (SOM)} (PROMETEO/2022/069).

\end{document}